\documentclass[twocolumn]{aastex62}

\newcommand\latex{La\TeX}

\shorttitle{The HH 212 Jet Complex}
\shortauthors{Reipurth et al.}

\begin{document}

\title{\large\bf   The Giant Herbig-Haro Flow HH~212 and Associated Star Formation           }

\correspondingauthor{Bo Reipurth}
\email{reipurth@hawaii.edu}

\author[0000-0001-8174-1932]{Bo Reipurth}
\affil{Institute for Astronomy, University of Hawaii at Manoa\\
640 N. Aohoku Place, Hilo, HI 96720, USA}

\author{  C.J. Davis }
\affiliation{  National Science Foundation, 2415 Eisenhower Ave, Alexandria, VA 22314, USA  }

\author{ John Bally  }
\affiliation{  CASA, University of Colorado, Boulder, CO, USA  }

\author{ A.C. Raga  }
\affiliation{ Instituto de Ciencias Nucleares, UNAM, M\'exico  }

\author{ B.P. Bowler  }
\affiliation{ McDonald Observatory and the Department of Astronomy, University of Texas at Austin, Austin, TX 78712, USA  }

\author{T.R. Geballe   }
\affiliation{  Gemini Observatory, 670 North Aohoku Place, Hilo HI 96720, USA }

\author{ Colin Aspin  }
\affiliation{ Institute for Astronomy, University of Hawaii at Manoa\\
640 N. Aohoku Place, Hilo, HI 96720, USA  }

\author{Hsin-Fang Chiang   }
\affiliation{ National Center for Supercomputing Applications, Urbana, IL 61801, USA  }


\begin{abstract}

The bipolar jet HH~212, among the finest collimated jets
  known, has so far been detected only in near-infrared H$_2$
  emission. Here we present deep optical images that show two of the
  major bow shocks weakly detected in optical [SII] emission, as
  expected for a bona fide Herbig-Haro jet. We present widefield H$_2$
  images which reveal two more bow shocks located symmetrically around
  the source and along the main jet axis. Additionally, examination of
  Spitzer 4.5~$\mu$m images reveals yet another bright bow shock
  further to the north along the jet axis; no corresponding bow shock
  is seen to the south. In total, the HH~212 flow has an extent of
  1050~arcsec, corresponding to a projected dimension of
  2.0~pc. HH~212 thus joins the growing group of parsec-scale
  Herbig-Haro jets. Proper motion measurements indicate a velocity of
  about 170~km/sec, highly symmetric around the source, with an
  uncertainty of $\sim$30~km/sec, suggesting a probable age of the
  giant HH~212 flow of about 7000~yr. The jet is driven by a deeply
  embedded source, known as IRAS~05413--0104.  We draw attention to a
  Spitzer near- and mid-infrared source, which we call IRS-B, located
  only 7\arcsec~ from the driving source, towards the outskirts of the
  dense cloud core. Infrared photometry and spectroscopy suggests that
  IRS-B is a K-type star with a substantial infrared excess, except
  that for an extinction of A$_V$ = 44 the star would have only a weak
  infrared excess, and so in principle it could be a K-giant at a
  distance of about 2~kpc.

\end{abstract}


\section{Introduction} \label{sec:intro}

Bipolar outflows are fundamental characteristics of young stars, and
have been detected at all evolutionary stages from Class~0 to
Class~III. These outflows are detected in molecular, atomic, and
ionized transitions, representing physical conditions from
shock-heated fast-moving plasmas to slower and denser molecular gas
entrained from the ambient medium. Outflows are important sources of
feedback into the ambient medium in star forming regions and form part
of the self-regulation of star formation. The most dramatic
manifestations of the outflow phenomenon are the Herbig-Haro jets,
highly collimated high-velocity bipolar flows that emit mainly in
Balmer lines, forbidden lines of [SII], [OII], [FeII], and molecular
transitions of H$_2$ (for reviews see e.g. Reipurth \& Bally 2001, Frank et al. 2014, and
Bally 2016). 

One of the finest collimated and highly symmetric jets known is
HH~212, discovered by Zinnecker et al. (1998). It is located in the
L1630 cloud complex in the tenuous 'gap' between the rich star forming
regions NGC~2023/2024 to the southwest and NGC 2068/2071 to the
northeast. Its distance is assumed to be $\sim$400~pc (Anthony-Twarog
1982, Kounkel et al. 2017). The closest region of star formation to
HH~212 is the Ori~B9 region to the southwest (see Figure~9 of
Miettinen 2012). The driving source of HH~212 is IRAS~05413--0104, a
Class~0 source embedded in a cloud core that is flattened and rotating
perpendicular to the jet flow axis (Wiseman et al. 2001, Lee et al.
2006, 2014). The HH 212 jet has been observed in the infrared in the
principal H$_2$ transition at 2.212~$\mu$m by Davis et al. (2000),
Smith et al. (2007), and Correia et al. (2009).  From radial
velocities and proper motions, the water masers are found to move
along the jet axis, and Claussen et al. (1998) deduced that the
outflow axis lies within 5$^\circ$ of the plane of the sky.  The jet
was studied in SiO by Takami et al. (2006), Codella et al. (2007), and
Cabrit et al. (2007, 2012), and the associated molecular outflow has
been studied in detail by Lee et al. (2000, 2006, 2007, 2008, 2015),
Codella et al. (2014, 2016) and Leurini et al. (2016).  ALMA
observations have revealed an accretion disk (Bianchi et al. 2017, Lee
et al. 2017, 2018a) and have been used to explore the jet launching
region and an associated disk wind (Podio et al. 2015, Codella et
al. 2016, Tabone et al. 2017, Sahu et al. 2018, Lee et al. 2018b).

The HH~212 driving source was detected in the millimeter continuum by
Zinnecker et al. (1992) and Chini et al. (1997), and in the radio
continuum at the VLA by Galv\'an-Madrid et al. (2004). Recently Chen
et al. (2013) found, using 1.3mm and 850~$\mu$m interferometry, that
it is a binary source, MMS-1 and 2, with a separation of
0.53$\pm$0.05\arcsec, and they detected a very faint third source,
MMS-3 at a separation of 1.2$\pm$0.1\arcsec, and coincident with a
source tentatively detected earlier at 1.4~mm by Codella et al.
(2007). However, these two additional sources have not been detected
in ALMA studies of the region, so the potential multiplicity of the
source needs to be examined further.

In this paper, we show that parts of the jet are observable at optical
wavelengths, and using widefield H$_2$ imaging we find new bow shocks
at larger distances from the driving source, demonstrating that the
HH~212 flow joins the ranks of the parsec-scale outflows. While the
HH~212 star forming event is often seen as a case of isolated star
formation, we also discuss evidence that several other young stars
have formed in the immediate vicinity of the jet source.  We also
discuss the nature of a near-infrared source, located 7\arcsec~ from
the driving source of HH~212, at a projected separation of 3000~AU.

\section{Observations} 
The HH 212 region was observed with infrared imaging and photometry at
UKIRT and Subaru, with infrared spectroscopy at Gemini, and with
optical imaging at the ESO New Technology Telescope, as described
below.


HH~212~IRS-B was observed with GNIRS (Elias et al. 2006) at the
Gemini-North 8m telescope on UT Jan 1, 2013 (program GN-2012B-Q-86)
with a total on-source exposure time of 2~hr 40~min under photometric
conditions. A 0.45 arcsec wide slit was used. Because of the nearby
jet, the slit was not positioned in the parallactic angle, and the
standard star, observed at the same airmass, was observed at the
same position angle (23$^\circ$).


The central part of HH 212 was observed at the 8m Subaru telescope on UT Jan 10, 2012 using
IRCS (Tokunaga et al. 1998) and H (1.64~$\mu$m) and K$'$ (2.15~$\mu$m)
filters of the Mauna Kea Observatories system (Tokunaga et al. 2002).
The conditions were photometric with a seeing around 0.9\arcsec~ at K$'$.
The image scale is 52 mas per pixel and the field of view is
54\arcsec~on a side.  The total integration time was 200 and 50
seconds in the $H$ and $K^\prime$ filters, respectively.  A 10-point
dithering was used to map the region.  The airmass during the
observations was around 1.1.  The data reduction was done using the
Image Reduction and Analysis Facility (IRAF), following standard
procedures described in the IRCS cookbook.  Following Connelley et al.
(2008), aperture photometry was performed using IMEXAMINE in IRAF with
five aperture sizes (0.9\arcsec, 1.2\arcsec, 1.5\arcsec, 1.8\arcsec,
and 2.1\arcsec).  The standard stars were measured in the same way and
the magnitudes from the UKIRT Faint Standards (Leggett et al. 2006)
were adopted to give magnitude estimates of the targets.  The $K$
magnitude from the standard catalogue is used as the standard $K'$
magnitude.  This is an acceptable approximation because the $K$ and
$K'$ filters have similar bandwidth in the Mauna Kea Observatories
Near-Infrared filter set.  Also, the standard stars used are early A
type stars which further mitigate the difference.  Measurements using
different aperture sizes and different standard stars are averaged and
errors are estimated.  Furthermore, a minimum photometric error is
estimated by the standard deviation of the standard star photometry.
For each target, the error is determined by the standard deviation of
individual measurements using different standard stars and different
aperture sizes.  The two errors are combined by a Pythagorean sum to
represent the overall photometric uncertainty.  For airmass
correction, we use the median extinction values from Krisciunas et al.
(1987).  The extinction coefficient value for the $K$-band is used for
our $K'$ observations.  The error introduced by this is of the order
of ~0.001 mag in terms of airmass correction, and much smaller than
the quoted errors.


HH 212 was observed with EMMI at the ESO New Technology Telescope
(NTT) on February 6, 1995 using a [SII] filter (ESO \#655). The
exposure time was 1800 sec.


A widefield near-infrared image of HH 212 was taken on February 14,
2010 through an H$_2$ filter ($\lambda \sim 2.1215$ micron,
$\delta\lambda \sim 0.021$ micron) using the UKIRT Wide Field Camera
WFCAM (Casali et al. 2007) in a seeing of 1.0 arcsec.  Individual 60
sec exposures were repeated with a 5-point dither pattern and 2$\times$2
micro-stepping (this yields a 0.2 arcsec pixel scale), resulting in a
total integration time of 1200~sec.  Comparison images were taken
through a K-band filter ($\lambda \sim 2.20$ micron, $\delta\lambda
\sim 0.34$ micron) on March 15, 2014 using the same camera, telescope
and dither/microstep procedure, resulting in a total integration time
of 200 sec. The seeing varied between 1.1 and 1.5 arcsec.

A second-epoch H2 image was obtained on January 5, 2018 also using
WFCAM on UKIRT in DirectorÕs Discretionary Time. On this occasion 40
sec exposures were repeated with a 9-point jitter and 2x2 micro-step
patter, resulting in a total integration time of 1440 sec. The seeing
was about 0.9 arcsec on this occasion.

Reduction of all of the WFCAM data was performed by the Cambridge
Astronomical Survey Unit (CASU).  The pipeline processing and science
archive are described in Irwin et al. (2004).

\section{Results} \label{sec:whatever}

\subsection{HH 212 Morphology from Infrared and Optical Images}

\begin{figure*}
\epsscale{1.0}
\plotone{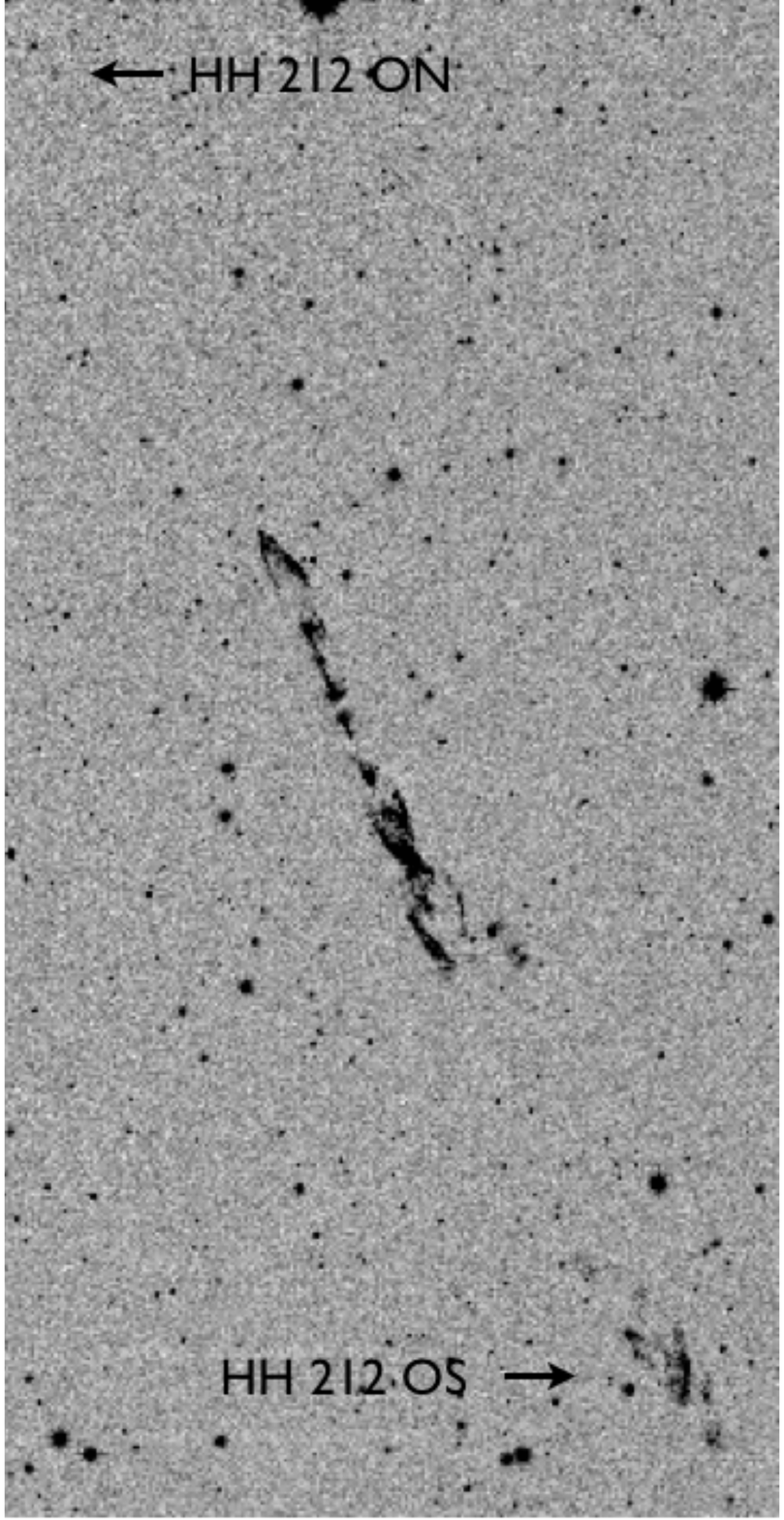}
\epsscale{0.8}
\caption{\em A molecular hydrogen image of the parsec-scale HH 212 flow 
obtained with WFCAM at UKIRT. North is up and east is left, and the 
vertical extent of the image is about 12 arcminutes. 
  \label{wide}}
\end{figure*}

The HH 212 jet known up to now is about 4~arcminutes in extent,
corresponding to a projected length of 0.46~pc at the distance of
400~pc, and is well studied on those scales.  However, many HH flows
have much larger dimensions (e.g., Reipurth et al. 1997), and we have
searched for further components along the flow axis.
Figure~\ref{wide} shows an image from the MHO Catalogue\footnote{The
MHO Catalogue can be found at
  http://www.astro.ljmu.ac.uk/MHCat, where HH~212 is listed as MHO~499
  (Davis et al. 2010).} obtained in the H$_2$ transition at
2.212~$\mu$m. To the south, approximately along the well defined flow
axis, a newly discovered large bow shock is seen, here labelled OS
(for Outer-South)\footnote{This has independently been discovered
by Mark McCaughrean (priv. comm.)}, it is very similar in morphology
and size to the two major shocks at the ends of the hitherto known
parts of HH~212. There is a slight change in orientation of about
4$^\circ$ in the line from the source to the apex of this new distant
bow shock relative to the axis of the innermost jet knots.  In fact,
close examination shows a gradual change in orientation from the
innermost to the outermost knots, suggesting precession of the flow
axis. The displacement is towards the west.

We have also searched north along the flow axis, and here find a very
faint diffuse knot along the precise same line through the source
region and the new OS bow shock, but slightly closer to the source.
The knot, which we label ON (for Outer-North), is very faint, and we
have verified that it is pure H$_2$ emission by comparing to a
broadband K-image. The displacement for this knot relative to the axis
of the inner jet knots is to the east, as expected for a precessing
flow axis. With these new shocks, the total projected extent of the HH
212 flow is now 12.6 arcminutes, which at 400~pc corresponds to
1.47~pc, so HH~212 joins the group of parsec-scale HH flows.

\begin{figure*}
\epsscale{1.0}
\plotone{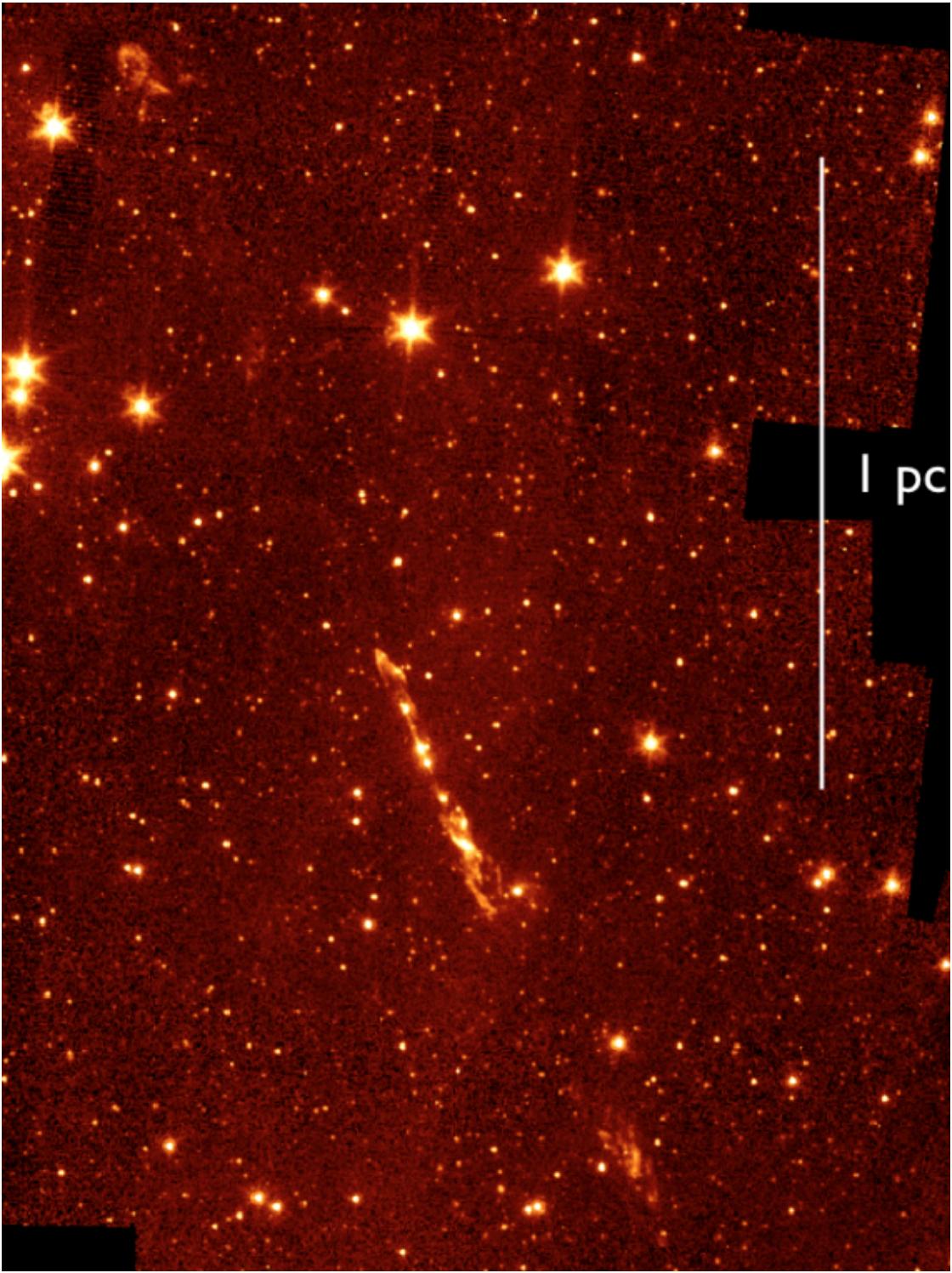}
\epsscale{1.0}
\caption{\em A Spitzer  wide field Channel~2 image (4.5 $\mu$m) of the HH 212 flow. All near-infrared components are seen, as well as a large bow shock outside the near-infrared WFCAM field. The total extent of the HH~212 is thus 1050~arcsec, corresponding to projected dimensions of 2.0~pc. 
\label{spitzer}}
\end{figure*}

Finally, we have examined the available Spitzer IRAC-2 (4.5~$\mu$m)
images obtained of this region. It is well known that many outflows
emit in the 4.5~$\mu$m band, which often is depicted as green in
multi-filter IRAC color images, and hence are dubbed 'Extended Green
Objects' or EGOs (e.g. Cyganowski et al. 2008).  Such objects,
including the HH~212 jet, are well seen in the Spitzer~4.5~$\mu$m
image, and the emission is believed to be mainly from the (0-0) S(9)
line of H$_2$ at 4.69~$\mu$m, possibly with additional CO vibrational
emission (Takami et al. 2010).
We have found yet another even more distant bow shock along the
northern outflow lobe, see Figure~\ref{spitzer}. No corresponding bow
shock is seen to the south. We dub this northern bow shock ON2. The
total distance from ON2 to OS is 1050~arcsec, which at 400 pc
corresponds to a projected separation of 2.0~pc. Takami et al. (2010)
studied the inner region of the HH 212 jet in the Spitzer 4.5~$\mu$m
and 8.0~$\mu$m filters, and showed an excellent correspondence with
H$_2$. Thus, although the bright and large bow shock ON2 lies outside
our UKIRT WFCAM field, we expect it to be emitting in H$_2$ (1-0) S(1)
rovibrational emission at 2.12~$\mu$m as well.

Figure~\ref{ir-opt}a shows an enlargement from the H$_2$ image,
presenting the well studied inner parts of the HH~212 jet.  The entire
jet emits in molecular hydrogen, but since Herbig-Haro objects are
defined in terms of optical emission lines (primarily H$\alpha$ and
[SII]), upon its discovery a [SII] image was obtained by one of us (BR)
with the ESO 3.5m NTT, where it was found that both of the main bow
shocks of HH~212 are faintly visible through the screen of extinction
that covers the jet. On this basis the jet complex was given the name
HH~212. A new deeper optical image from the Subaru telescope is shown
in Figure~\ref{ir-opt}b, where the faint visible HH knots are
marked with circles.


\subsection{Proper Motions}

Proper motions were derived from our 2010 and 2018 UKIRT images
obtained through an H$_2$ filter. The time difference is 2882
days. During this time the jet moved measurably, and we have used a
wavelet convolution method to derive the proper motions; the method is
described in Raga et al. (2017).  Figure~\ref{propermotions} shows the
resulting velocity vectors and the values in km/sec are listed in
Table~2. The motion along the jet axis is evident. While the error box
on each individual vector is non-negligible, taken together the motion
is well established. The X and Y values in Table~2 were measured with
the jet oriented along the Y-axis. Given the very well defined flow
axis, we interpret the V(X) values perpendicular to the jet axis as
the uncertainty of the measurements, yielding a value of 30 km/sec on
individual velocities.  The values along the jet-axis are assumed to
be due to the physical motion of the knots, and we find a remarkable
symmetry around the source, with identical values of the northern
(171~km/sec) and southern jet (170~km/sec). Since the jet is oriented
only about 5$^\circ$ from the plane of the sky this projected velocity
also corresponds to the space motion of the jet. The motion of the new
southern bow shock is around 180 km/sec, albeit with a much larger
spread in V(X) of 50 km/sec. However, the result suggests that the jet
is not slowing down as it ploughs through the ambient medium, in
contrast to the HH~34 giant jet complex, which shows a major
deceleration from source to terminal shocks (Devine et al. 1997). The
new most distant bow shock (ON2) is located outside our UKIRT images,
and hence we have not obtained its proper motion.

Assuming a constant velocity of $\sim$170~km/sec and a distance of
400~pc, the innermost pair of bright knots (NK1 and SK1
) have ages of
75~yr, and the next pair of bright knots (NB1/2 and SB1/2) are 500~yr
old. The last symmetric pair of bright shocks is NB3 and SB3, which
are 1050~yr old. After that we see three more distant bright bow
shocks, SB4 at 1550~yr, the new southern bow shock OS at 4300~yr and
the most distant northern bow shock ON2 at 7000~yr. We thus see a
progression of larger and larger intervals between the bow shocks as
we move away from the source, and this is discussed further in
Section~4.

\subsection{Jet Opening Angle}

Jets gradually expand as they leave their sources, and it has been
shown that close to the source they expand sideways very rapidly,
after which the expansion slows down and reaches a constant rate
(e.g., Reipurth et al. 2000).

The opening angles of the finely collimated jets HH~1 and HH~34 have
been measured accurately on HST images, and the half-opening angle for
HH~1 is 1.3$^\circ$ and for HH~34 it is 0.4$^\circ$ (Reipurth et
al. 2000, 2002). We have used archival HST NICMOS
images\footnote{Program 7368, PI: M. McCaughrean} to measure the
opening angle for the HH~212 jet. Figure~\ref{openingangle} shows the
inner 30~arcsec of the jet around the source. Due to high
obscuration, the innermost part of the jet is not visible even in
these near-infrared images. At their widest, the well resolved bright
inner knot pair NK1 and SK1 (using the nomenclature of Zinnecker et
al. 1998), has a mean width of 1.56$\arcsec$. This leads to a mean
half-opening angle for this knot pair of 6.8$\pm$0.5$^\circ$. With an
approximate tangential velocity of 170~km~s$^{-1}$ the two knots were
ejected from the source 73 yr ago. Under the assumption that the jet
started within the innermost few AU of the star-disk system, this
implies that the two knots have a mean radial expansion velocity of
20~km~s$^{-1}$. In the internal working surface model of Raga et
al. (1990), the jet knots are small working surfaces moving into a
co-moving medium. The morphology of the knots NK1 and SK1 very much
fit that interpretation. A working surface consists of an outer bow
shock and an inner jet shock or Mach disk. In a rest frame moving with
these two shocks, ambient material moves into the bow shock and jet
gas moves into the Mach disk, leading to gas being expelled sideways
and a consequent lateral expansion of the knot. It is difficult to
compare this measurement with shock models, because the sideways
ejection velocity is a function of several poorly know parameters,
especially the cooling rate.  Given that molecular hydrogen is highly
excited in HH~212, and the S(1) line of H$_2$ that we observe in
HH~212 is excited in weak shocks, it follows that the lateral gas
expulsion in the NK1/SK1 knots may be highly supersonic.

The opening angle of the HH~212 jet is significantly larger than
measured for the HH~1 and 2 jets. One possible explanation for this
may be that the HH~212 shocks are stronger, leading to a faster
lateral expansion of the bow shock wings. We have also measured the
width of the next two knots, NK2 and SK2, and find a half-opening
angle of 2.3$^\circ$, which is significantly less than for the NK1/SK1
pair, but still larger than for the HH~1 and HH~34 jets.




\subsection{A New Wide Companion?}

While studying the HH 212 jet on Spitzer IRAC images, we noticed a
nearby source with a strong 8~$\mu$m flux, unlike any of the other
numerous background stars in the region. We have studied this source
further, and will below argue that this is possibly a new distant
component, here called IRS-B, in the Class~0 multiple system IRAS
05413---0104, or HH~212~MMS.

\subsubsection{Photometry}

Figure~\ref{triglyph}a,b shows images obtained in H- and K-filters of the
innermost region of the HH~212 jet. In addition to multiple knots in
the jet, three near-infrared sources are seen. IRS-B is marked, located at
 $\alpha_{2000}$ 05:43:51.2, $\delta_{2000}$ --01:02:47. 
 The MMS source was detected at the VLA by Galv\'an-Madrid et al.
 (2004) at $\alpha_{2000}$ 05:43:51.408 $\delta_{2000}$ --01:02:53.13.
 The projected separation of MMS and IRS-B is thus 6.9~arcsec. Assuming
 that a line between the two sources lies at an angle of 30$^\circ$ to
 the plane of the sky, then at the assumed distance of 400~pc the
 physical separation of the two sources is about 3000~AU.





We have imaged IRS-B with the IRCS at the Subaru 8m telescope in
H and K' filters. The star is readily detectable in both filters,
see Figure~\ref{triglyph}a,b, with H = 19.20$\pm$0.10 mag and K' =
15.95$\pm$0.11 mag.  IRS-B is also detected in all the four IRAC bands
of Spitzer, yielding [3.6] = 14.01$\pm$0.14, [4.5] = 13.16$\pm$0.12,
[5.8] = 12.67$\pm$0.12, [8] = 11.15$\pm$0.15~mag.

Figure~\ref{triglyph}c shows the IRAC 8~$\mu$m image, in which IRS-B
is clearly visible. IRS-B is not seen in the 24~$\mu$m MIPS image,
possibly due to its proximity to the HH~212 driving source, which is
also a likely reason it is not seen in the submm and mm images of Chen
et al. (2013).

Figure~\ref{BB} shows the available photometry of HH~212~IRS-B from
1.6 to 8 $\mu$m as black circles. Using the reddening curves of
Fitzpatrick (1999) with $A_V/E(B-V)$ = 3.1, the observed photometry
has been dereddened by an A$_V$ of 15 mag (blue triangles), 30 mag
(green squares) and 44 mag (red circles). Two blackbody curves of
4570~K and 3916~K, corresponding to K0 and M0 giants (van Belle et al.
1999) are shown fitted to the H-band fluxes; the arguments for
choosing this spectral range are based on a near-infrared spectrum and
are discussed in the next section.

If one plots the Spitzer IRAC photometry in color-color diagrams (e.g.,
Gutermuth et al. 2008), IRS-B falls in the region of Class~II sources
bordering on Class~I sources. A potentially high extinction, however,
complicates the interpretation, and we therefore proceed as follows.

An important question to ask is whether IRS-B has any infrared excess.
The answer is made difficult by the lack of detections shortwards of the
H-band. If we make the assumption that the shortest wavelength
detection in the H-band represents the photosphere of the star, then we
can fit the abovementioned black body curves at this wavelength. For
low extinctions, IRS-B has a major infrared excess at all observed
wavelengths. By increasing the assumed extinction to A$_V$ = 44 mag.,
it is possible to fit the H-band and the [3.6], [4.8], and [5.6] bands
to the black body curves, but both the K-band and [8]-band fluxes
indicate excess. If one increases the extinction to larger values
while maintaining the assumption that the H-band flux is photospheric,
then the IRAC band fluxes fall underneath the black body curves, which
is evidently unphysical. Higher extinctions than A$_V$ = 44 mag. are
possible, but then imply a near-infrared excess in both H- and
K-bands. We conclude that the available photometry implies that IRS-B
has an infrared excess, which may be significant for either low or very high
extinction, or more modest if the extinction is around A$_V$ = 44 mag.

If IRS-B is associated with the HH~212 star forming region, then its
luminosity would be roughly 0.5~L$_\odot$ if it follows the red curve
for A$_V$=44 mag in Figure~\ref{BB}, and less if the extinction is
lower. This is consistent with a young low-mass star.

If IRS-B is {\em not} a young companion to HH~212-MMS, then what is
it? Statistically the most likely background star would be a
K-giant. Assuming an extinction of A$_V$=44, and ignoring the small
observed infrared excess for this extinction value, we find that a
K-giant at a distance of roughly 2 kpc could match the photometry.
Unfortunately, IRS-B is not visible in the optical, so Gaia cannot
help in distinguishing between these possibilities.

It follows that on the basis of photometry alone, we cannot decide
conclusively whether IRS-B is young or not.







\subsubsection{Spectroscopy}

Figure~\ref{GNIRS} shows a near-infrared spectrum in the H- and
K-bands of HH~212~IRS-B. As expected from the near-infrared photometry
provided above, the spectrum is extremely red, with only a low S/N in
the H-band, and no signal at all in the J-band. In the K-band, a
photospheric spectrum is seen, with pronounced CO band-head absorption
between 2.3 and 2.4~$\mu$m, and weak Na~I, Ca~I, and Mg~I lines. Also,
a weak H$_2$ 2.122~$\mu$m emission is seen, but given the strong and
extended H$_2$ emission from the nearby jet, we cannot exclude that
this might result from incomplete sky subtraction.



To constrain the spectral type of HH 212 IRS-B, we compare the
2.13-2.43~$\mu$m region of our $K$-band spectrum to the IRTF library
of cool stars from Rayner (2009) and Cushing (2005).  This region
contains a wealth of absorption lines sensitive to both temperature
and surface gravity.  We make use of both dwarf templates and giant
templates for this exercise since this range of gravities likely
brackets that of IRS-B.  Our raw data are complicated by high
reddening and possible excess emission, so for this comparison we
first normalize both the science spectrum and IRTF templates by
fitting and subtracting a fourth-order polynomial to the continuum in
that region.  Following the method outlined in Bowler et al. (2009),
we optimally scale each template to HH 212 IRS-B by computing the
scale factor that minimizes the resulting $\chi^2$ value.
The reduced $\chi^2$ values exhibit broad minima between
$\approx$G3 and K7 spectral types for both the giant and dwarf spectra.



Closer examination of the spectrum in Figure~\ref{GNIRS} suggests a
spectral type of IRS-B between K0 and M0. The absence of any water
vapor bands in the GNIRS spectrum suggests that a later spectral type
than early M is unlikely.  The spectrum shows no sign of the lines of
Ca/Fe/Mg around 1.95-2.00~$\mu$m which become increasingly strong
through the K spectral type. Also, there is no trace of TiO bands,
which are dominant in M-stars.  Hence it appears that, due to the
extreme reddening, the spectral type of IRS-B cannot be determined
accurately, and can be only loosely constrained to between K0 and
M0, with K5 as the best fit.

In summary, the evidence points to IRS~B being a young star, but the
possibility that it could be a background K-giant cannot be excluded.

\subsection{An Edge-on Disk and a Class~I source}

While examining the UKIRT images, we noticed what appears to be an
edge-on disk 2~1/2 arcmin to the south-west of the HH212 source, see
Figure~\ref{edgeondisk}. This object, which is located at
$\alpha_{2000}$ 5:43:45.41, $\delta_{2000}$ -1:04:55.5, is not
detected by 2MASS nor by WISE, and we refer to it as IRS-C.
Reflection nebulae are apparent on either side of the disk, as well as
a central faint point-like object, which presumably is the
illuminating embedded star. The obscuration can be traced for about
10~arcsec on either side of the source, corresponding to a width of
about 8000~AU. This is far too large to be a physical disk, and is
more likely to represent the shadow cast by a much smaller
circumstellar disk on its surroundings (see Hodapp et al. 2004).
There is also a nebulous star about 18~arcsec to the NE of the edge-on
disk, identified as 2MASS J05434630-0104439 and classified as a young
star from Spitzer photometry (Megeath et al. 2012).  We here refer to
this object as IRS-D. The 2MASS colors are very red, and this
continues into the WISE wavelength range, with W1=12.67, W2=11.19,
W3=7.32, and W4=3.07, where it is the dominant source (see
Figure~\ref{WISE}). When placed in a [3.4]--[4.6] vs [4.6]--[12] WISE
diagram, it falls in the middle of the Class~I protostar region (see
Koenig et al. 2012). These two objects demonstrate that although the
HH~212 source is separate from the nearest star forming complex
Orion~B9, it is not a completely isolated star forming event.




\section{Discussion}

In the following we offer some speculations on the origin of IRS~B
assuming it is a young star.

Giant HH flows provide insight into the accretion history of their
driving sources. Reipurth (2000) suggested a scenario which accounts
for all the characteristics of giant outflows in terms of the
evolution of an unstable triple system.

The most common outcome of the collapse and fragmentation of a cloud
core is not the formation of a single star, but rather a binary or a
small multiple system. This has long been suspected on theoretical
grounds (Larson 1972), but is increasingly supported by observations
at many wavelengths (e.g., Chen et al. 2013; for reviews see Goodwin
et al.  2007 and Reipurth et al. 2014). When more than two stars are
bound together in a non-hierarchical fashion, their motions are
chaotic and inevitably result in the decay of the system, which either
disintegrates or is re-structured into a hierarchical configuration
(e.g., Valtonen \& Mikkola 1991).  This typically occurs during the
protostellar phase (Reipurth 2000), and numerical simulations show
that ejected components often for a while remain tenuously bound in
wide orbits. While protostars generally are deeply embedded, such
orphaned protostars, as they are dubbed, can be flung to the outskirts
or outside of their nascent cloud core, making them observable at
near-infrared or even at optical wavelengths (Reipurth et al. 2010).
The ejected member is normally the lowest-mass component, and if the
ejection occurs early enough that the stellar embryo has not
accumulated sufficient mass to eventually burn hydrogen, it will
remain a brown dwarf in the absence of further mass growth (Reipurth
\& Clarke 2001).  Observations have indeed shown that the fraction of
distant companions to Class~I protostellar sources decreases
dramatically as these sources become more evolved (Connelley et
al. 2008).

It is during close triple approaches that the stars can exchange
energy and momentum, and one of the bodies can be flung out. Such
close interactions lead to massive accretion and outflow events,
creating the large bow shocks seen at the extremities of giant HH
flows. The remaining binary will have a highly eccentric orbit, and so
when the two companions have their first periastron passage, their
disks collide and another accretion/outflow event takes place. Due to
such viscous interactions, the binary may slowly spiral in while
becoming less eccentric, causing further and more closely spaced bow
shocks.

In this picture, IRS-B would be an orphaned protostar residing at the
edge of the HH 212 cloud core. This requires that the driving source
of HH~212 is a binary. As noted earlier, the companion detected by
Chen et al. (2013) is in doubt. However, such a binary would be so close
that it would be hard to detect. If we assume that the hypothetical
binary has a total system mass of 1 M$_\odot$ and a period of 58
years, which is the time it would take the two innermost knots (NK1
and SK1 in the nomenclature of Lee et al. 2007) to reach the positions
of the next pair of knots (NK2 and SK2) then we can derive a binary
semimajor axis of 15~AU, which at the distance to HH 212 corresponds
to 0.04~arcsec, assuming no projection effects. Such a companion
would thus not be detected in currently available data.

If IRS-B is associated with the HH~212 region, then it is interesting
to ask whether it is bound to the putative MMS binary, or it is
escaping. And if it is bound, is it a stable or unstable system? In a
major set of numerical simulations of triple systems, Reipurth \&
Mikkola (2012) calculated the ratios of stable, unstable, and
disrupted triple systems as a function of projected separation. For a
projected separation of about 3000~AU and an age of 1~Myr, it was
found that only about 5\% were unbound, while about 25\% were stable
triples, and about 70\% were unstable triples. At first glance this
would seem to contradict the fact that about 50\% of all triple
systems break apart during the protostellar phase. This is certainly
true, but those escaping bodies have already moved much further away
from their protostellar brethren.  Only about 5\% of unbound
companions would still have a relatively small projected separation of
around 3000~AU.  While we are not able with current techniques to
observationally determine whether IRS-B is bound or not, on a
statistical basis we can state with some confidence that it is highly
likely that IRS-B is still bound to the MMS binary, but only
tenuously, and during one of the coming periastron passages, the
system is likely to break apart, releasing IRS-B into a slow, gentle
escape.

If IRS-B is indeed young, then the most interesting aspect of the star
is that it is at all observable at near-IR wavelengths, even though it
must have a very young age similar to the Class~0 source IRAS
05413-0104. Such newly born protostars are generally so deeply
embedded that they are not observable at short wavelengths, but here
we may have the opportunity to witness the stellar part of a very
young protostar, thanks to its ejection from the deep interior of its
cloud core.  Because of its possible extreme youth, IRS-B should be
located high up on its Hayashi track, and should therefore in
principle be significantly more luminous than later in its
evolution. However, if it was ejected due to stellar triple dynamics,
it would have lost some or much of its disk and envelope, leaving
mostly the star itself to produce its luminosity.  It will eventually
appear in the optical as a late-type T~Tauri star.

\section{Conclusions}

We have studied the region of the HH~212 bipolar
jet and have obtained the following results:

1. The HH~212 bipolar jet is highly obscured and is visible mainly in
   molecular hydrogen transitions, but deep optical images show that
   the northern and southern bow shocks are optically visible.

2. We have discovered new distant bow shocks along the well defined
   flow axis of HH~212, increasing the size of the jet complex to
   $\sim$2.0~pc and increasing the dynamical age of the system by a
   factor of more than 4.

3. We have measured the proper motions of the HH 212 jet complex, and
   for the inner highly symmetric pairs of knots we find a velocity of
   170~km/sec with an uncertainty of about 30~km/sec. The new southern
   giant bow shock moves with the same velocity, suggesting that the
   jet is not slowing down as the flow penetrates the ambient
   medium. Since the angle of the HH 212 flow to the plane of the sky
   is only 5$^{\circ}$ the above velocity then equals the space
   velocity of the flow. For this velocity, the age of the giant HH
   212 system is 7000~yr.

4. A Spitzer-detected near- and mid-infrared source has been noted
   7\arcsec~ from the HH~212 source. The source shows an infrared
   excess for all extinction values, but it is minimal for
   A$_V$=44 mag. A H- and K-band spectrum shows a highly reddened
   K-type spectrum. If a background star it would be a K-giant at a
   distance of about 2~kpc. But the near- and mid-infrared photometry
   is also consistent with an embedded Class~II source. We cannot
   distinguish between these two possibilities, but note that if IRS-B
   is associated with the HH~212 cloud core, then it is likely an
   orphaned protostar which -- due to three-body interactions -- has
   become detectable at near-infrared wavelengths. If so, the statistical
   analysis by Reipurth \& Mikkola (2012) indicates that there is only
   a 5\% chance that HH~212~IRS-B is currently escaping, but there is
   a 70\% chance that it is presently only so tenuously bound that it
   will break loose when it dives into the cloud core during one of
   its coming periastron passages around the central protostars.

5. The HH~212 star forming event is located in a tenuous area of the
   L1630 cloud where the nearest star formation is in the Ori B9
   region further to the southwest. We have shown that the HH~212
   source is not so isolated, but has at least one Class~I and one
   Class~II object adjacent to it.

\acknowledgments

We thank the referee, Pat Hartigan, for helpful suggestions, and Tom Megeath for comments on the paper.
We are grateful to Watson Varricatt and Tom Kerr for obtaining the
UKIRT second-epoch image of HH 212 through Director's Discretionary
Time. UKIRT is owned by the University of Hawaii (UH) and operated by
the UH Institute for Astronomy; operations are enabled through the
cooperation of the East Asian Observatory. When some of the data
reported here were acquired, UKIRT was operated by the Joint Astronomy
Centre on behalf of the Science and Technology Facilities Council of
the U.K.
This project was supported by the Gemini Observatory, which is
operated by the Association of Universities for Research in
Astronomy, Inc., on behalf of the international Gemini partnership of
Argentina, Brazil, Canada, Chile, and the US.
Based in part on data collected at the Subaru Telescope, which is operated by
the National Astronomical Observatory of Japan (NAOJ).
and on observations collected with the NTT at the European Organisation for 
Astronomical Research in the Southern Hemisphere. 
We acknowledge use of IRAC and MIPS data from the Spitzer Science Center.  
This research has made use of the SIMBAD database, operated at CDS,
Strasbourg, France, and of NASA's Astrophysics Data System
Bibliographic Services,
and of ESASky, developed by the ESAC Science Data Centre (ESDC) team and maintained alongside other ESA science mission's archives at ESA's European Space Astronomy Centre (ESAC, Madrid, Spain).
This material is based upon work supported by the National Aeronautics
 and Space Administration through the NASA Astrobiology Institute under
 Cooperative Agreement No. NNA09DA77A issued through the Office of Space
 Science.

\vspace{5mm}
\facilities{UKIRT(WFCAM), Gemini-North(GNIRS), Subaru(IRCS), ESO-NTT(EMMI)}

\clearpage

\clearpage

\vspace{-2cm}


\begin{figure*}[t]
\epsscale{1.1}
\plotone{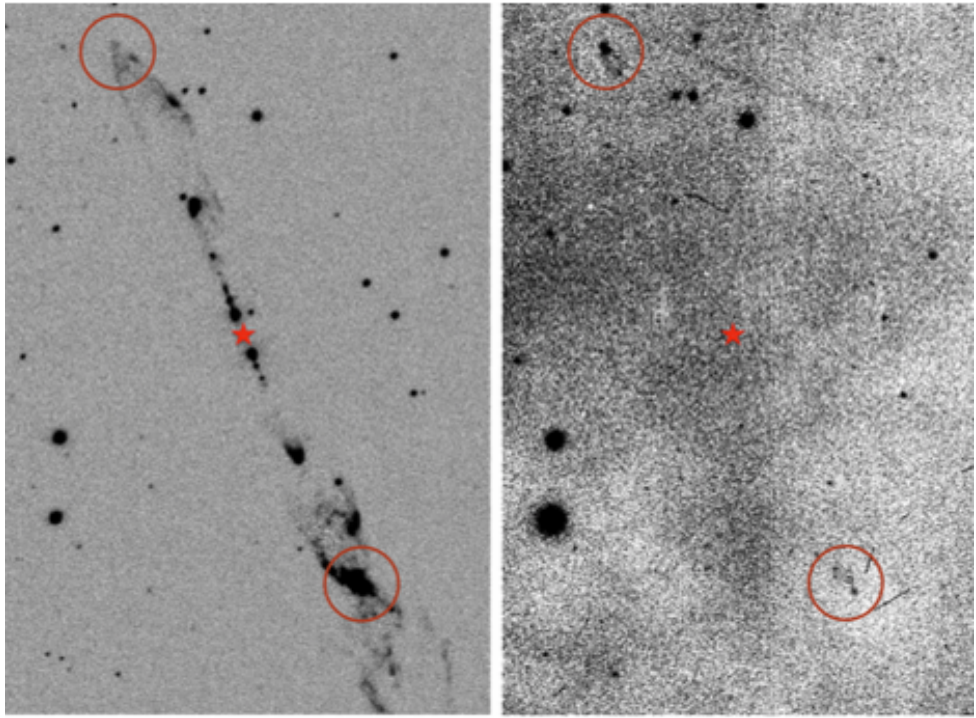}
\vspace{-5cm}
\caption{\em (a): H$_2$ 2.12 $\mu$m image of the HH 212 jet from
UKIRT. The embedded Class~0 driving source is marked with an asterisk.
(b): [SII] image of the approximate same region from the Subaru 8m
telescope; the faint optical HH objects are marked by circles. Both
images have north up and east left, and the vertical extent is about
4.5 arcminutes.  \label{ir-opt}}
\end{figure*}



\begin{figure*}[b]
\epsscale{1.1}
\plotone{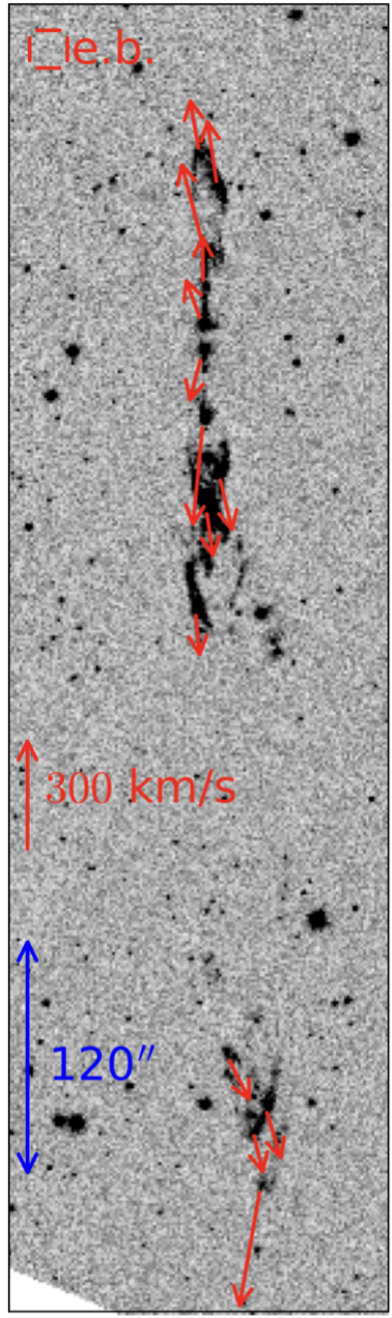}
\vspace{-0.5cm}
\caption{\em Proper motions of the HH 212 knots. A typical error box is indicated in the upper left corner.
\label{propermotions}}
\end{figure*}

\clearpage

\begin{figure*}
\epsscale{0.42}
\plotone{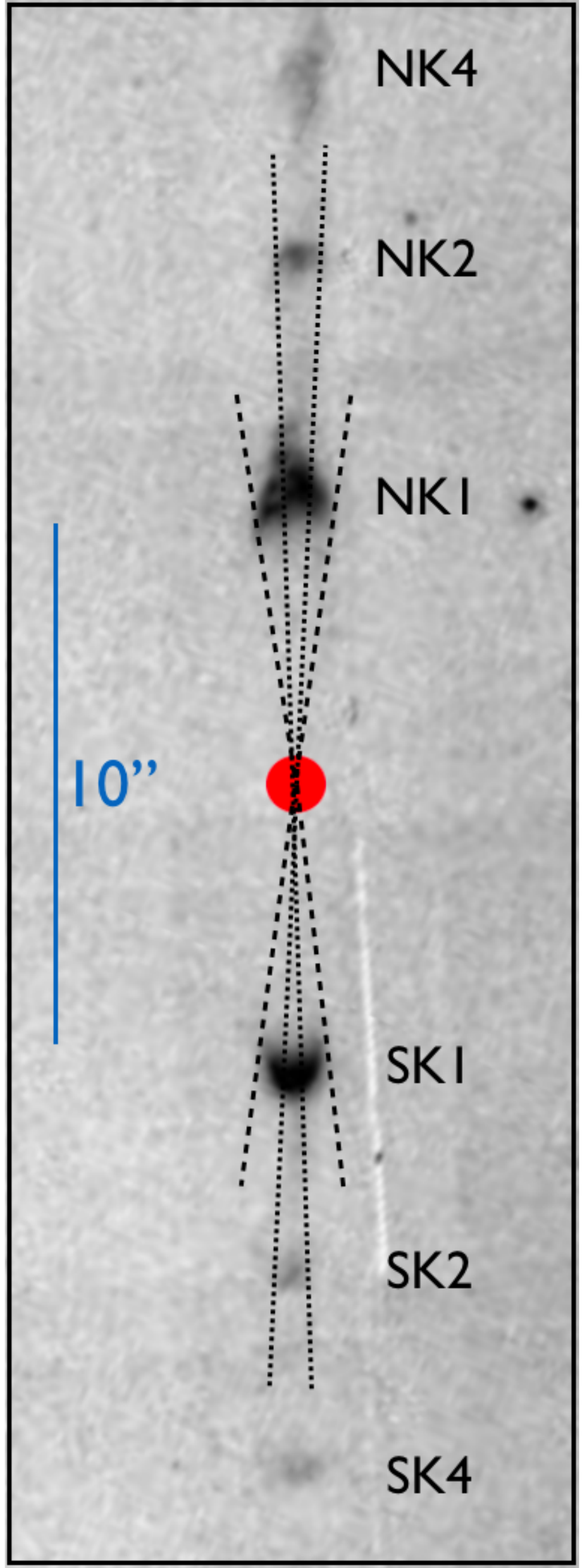}
\epsscale{0.9}
\caption{\em The innermost part of the HH~212 jet is seen here in an HST NICMOS H$_2$ image. The opening angle of the jet as measured for the first and the second pairs of knots is indicated. 
\label{openingangle}}

\vspace{-4.0cm}
\epsscale{0.7}
\plotone{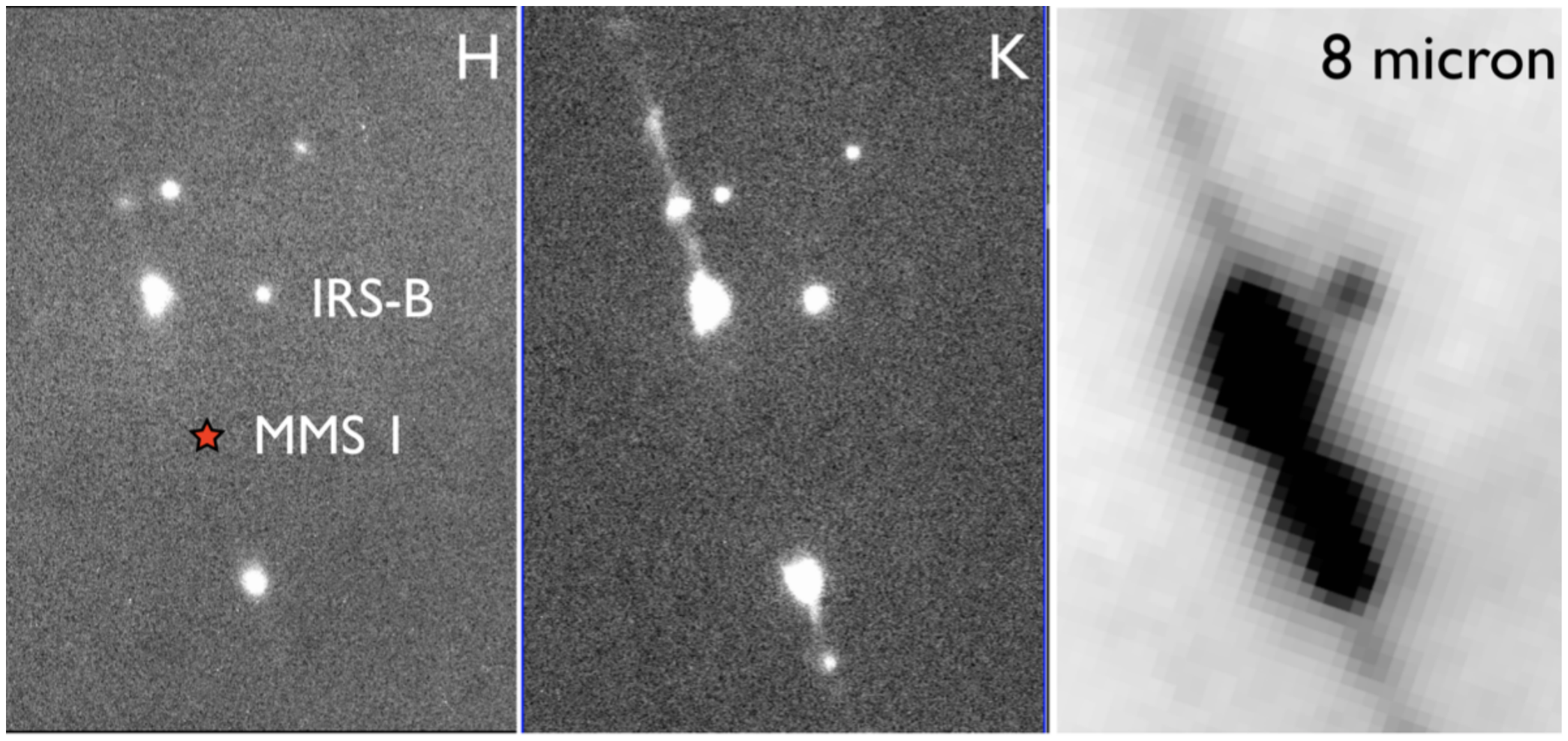}
\vspace{-4.5cm}
\caption{\em A mosaic of H- and K-band images of the center of the HH 212 
flow taken with the Subaru 8m telescope, and an 8~$\mu$m image from Spitzer.
IRS-B is marked, 7\arcsec~ from the location of MMS. 
[fig:HKimage]  
North is up and east is left. \label{triglyph}}

\vspace{-0.5cm}

\epsscale{0.37}
\plotone{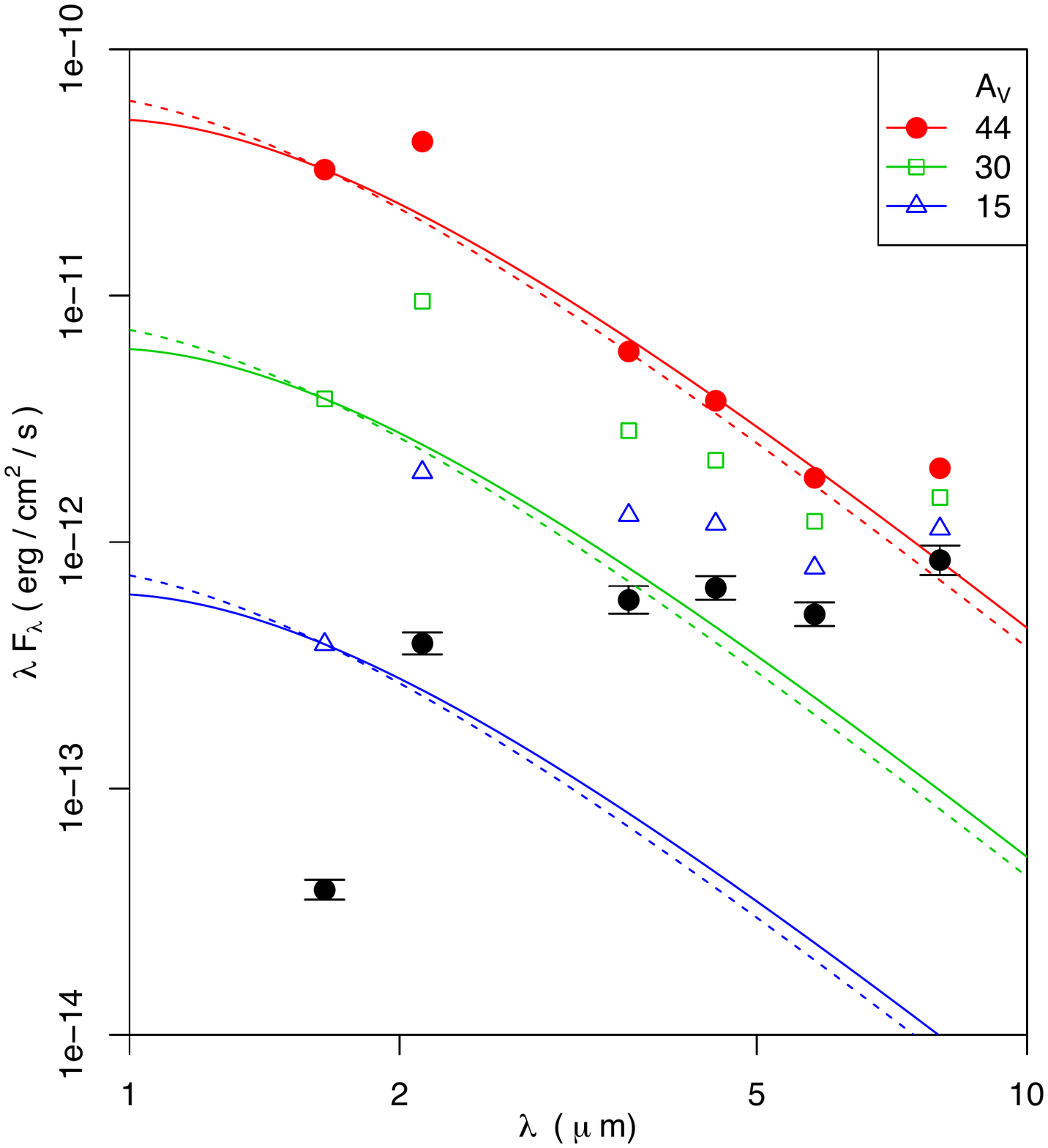}
\vspace{-1cm}
\caption{\em Available photometry of HH~212~IRS-B from 1.6 to 8~$\mu$m
  is plotted as black circles, the surrounding lines indicate the
  photometric uncertainty. The blue triangles, green squares and red
  circles show the same photometry de-reddened by A$_V$ = 15, 30, and
  44 magnitudes. Two black body curves for 3916~K (solid curve) and
  4570~K (dashed curve) are fitted to the shortest wavelength
  photometric point for all three de-reddened data sets; these
  temperatures correspond to M0 and K0 giants.
\label{BB}}
\end{figure*}

\clearpage

\begin{figure*}
\epsscale{0.7}
\plotone{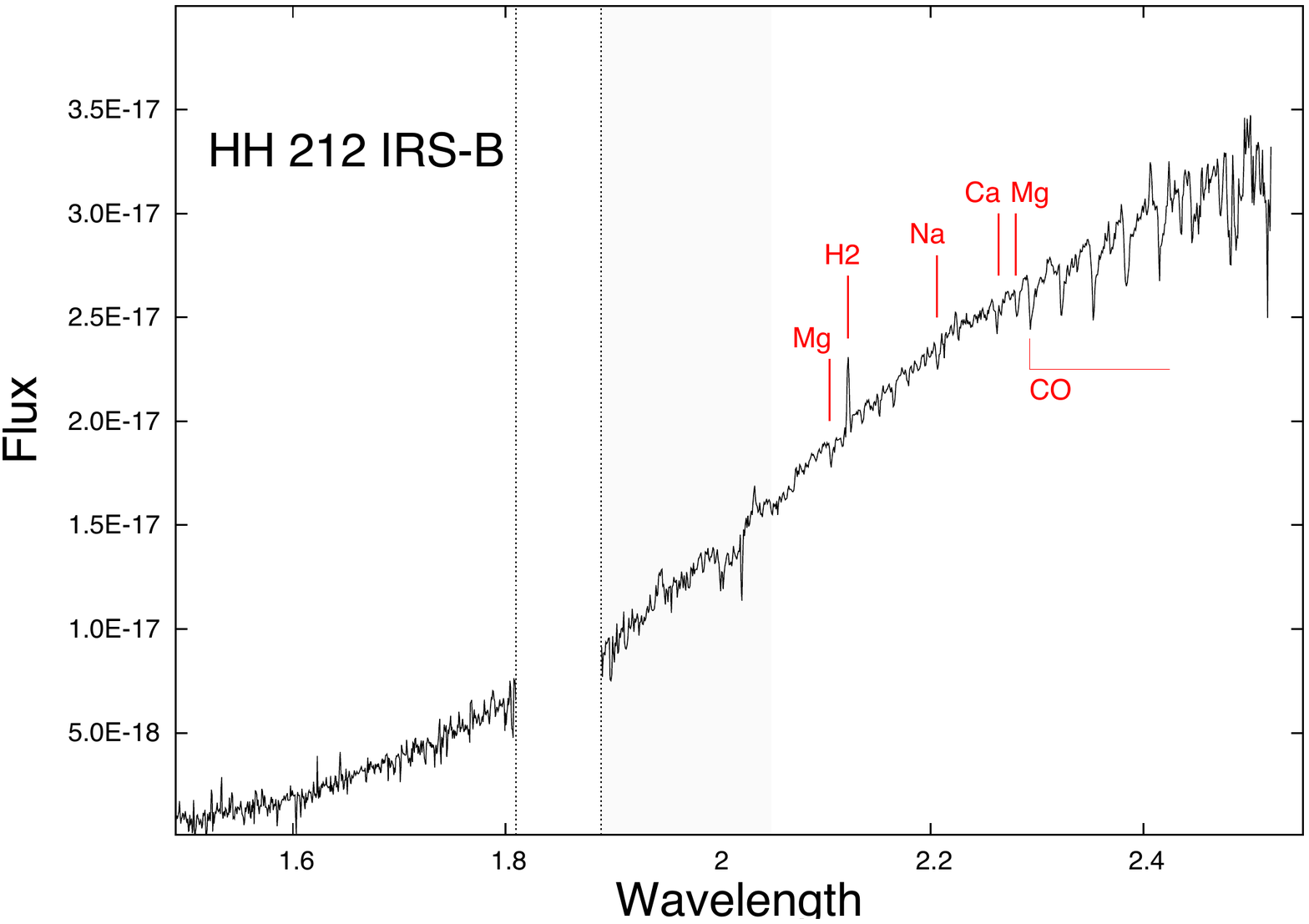}
\vspace{-3cm}
\caption{\em A GNIRS spectrum of HH 212 IRS-B from Gemini-North. Spectral 
features are indicated. The spectral region from 1.81 to 1.89~$\mu$m 
(between dotted lines) with strong atmospheric absorption (transmission 
$<$20\%) has been removed, while the region from 1.89 to 2.05 $\mu$m with
moderate atmospheric absorption (transmission $<$80\%) is indicated by 
grey shading. 
\label{GNIRS}}

\vspace{-4cm}
\begin{center}
\includegraphics[angle=0,width=13cm]{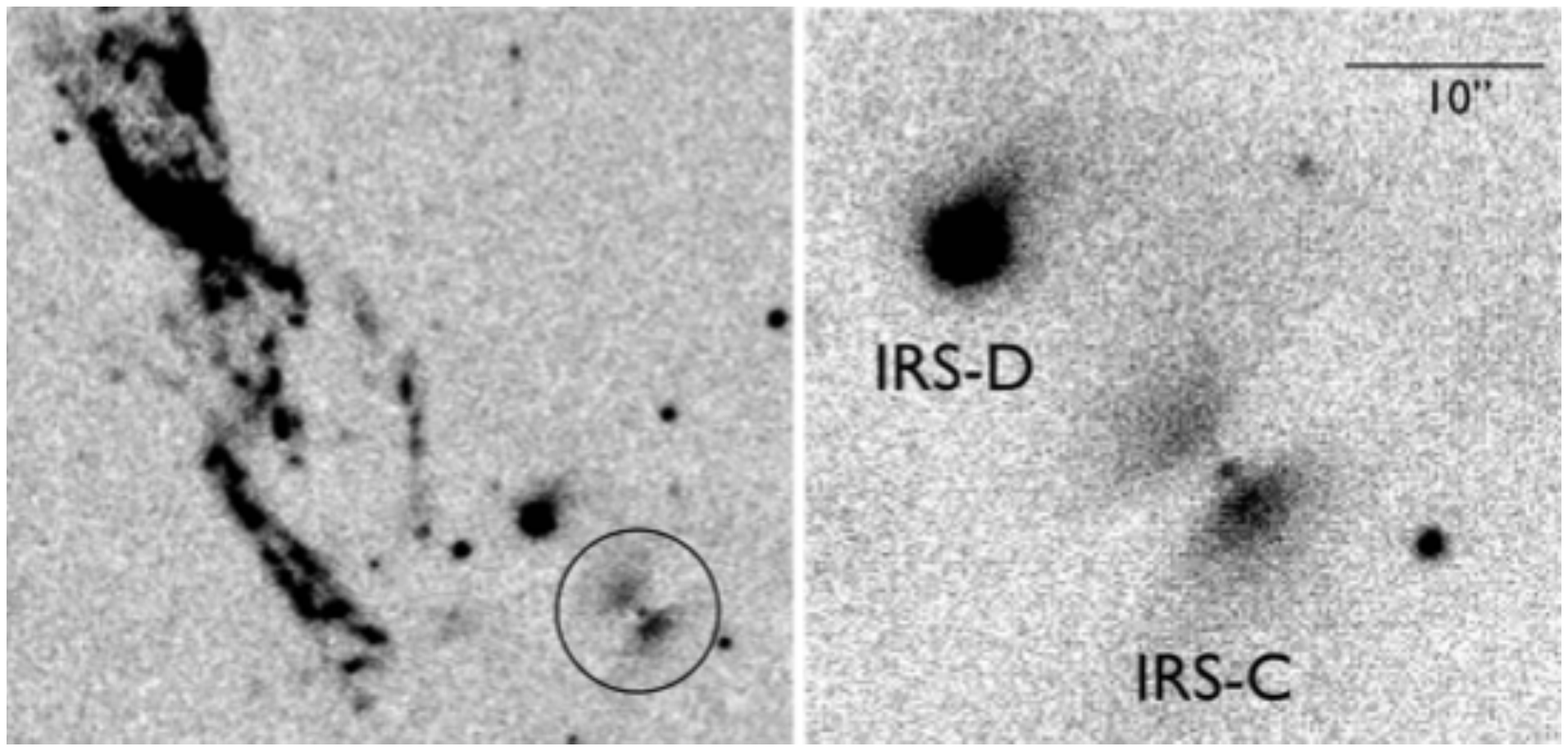}
\vspace{-5.5cm}
\caption{\em (a): An enlargement from the molecular hydrogen image in 
Figure~\ref{ir-opt}a showing a small edge-on disk (encircled). The
central star is faintly detected, and here called IRS-C. This new
source is not detected by 2MASS or WISE, but is bright in a
Herschel~250~$\mu$m image.
(b) A K-band image of a region $\sim$38$\times$40~arcsec showing the 
edge-on disk and its illuminating source IRS~C. The star 18~arcsec to
the NE is nebulous, and so apparently also young. It is here called
IRS-D. North is up and east is left.
 \label{edgeondisk}}
\end{center}
\end{figure*}

\clearpage

\begin{figure*}
\vspace{-1.5cm}
\begin{center}
\includegraphics[angle=0,width=15cm]{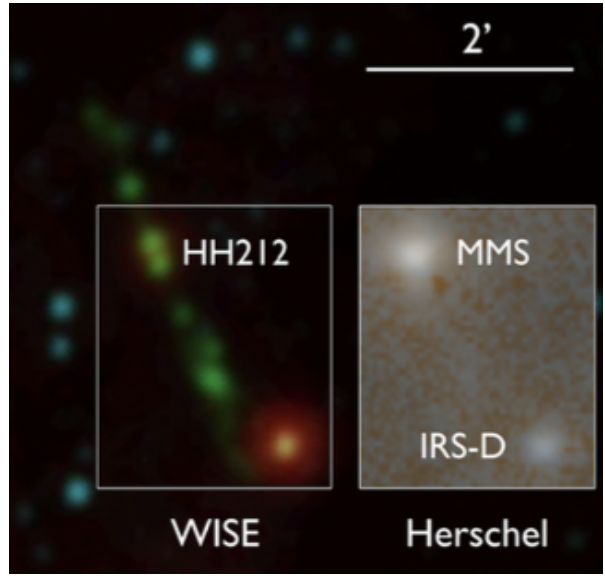}
\vspace{-5.8cm}
\caption{\em The HH~212 jet is seen in this WISE color image (blue 3.4~$\mu$m, green 4.6~$\mu$m, red 22~$\mu$m) together with the 2MASS source J05434630-0104439 (IRS-D) which is very bright at mid-infrared wavelengths. 
The framed region contains the two protostars MMS and IRS-D, and the
insert to the right shows the same region as observed by Herschel at
70/160~$\mu$m. It is seen that the Class~0 source MMS is only visible
at far-infrared wavelengths with Herschel, where it dominates the
Class~I source IRS-D that in turn dominates at near- and mid-infrared
wavelengths. North is up and east is left.
 \label{ir-opt}}
\end{center}
\end{figure*}


\begin{deluxetable}{lrrl}
  \tablewidth{0pt}
  
\centering

\tablecaption{Positions of New Bow Shocks$^a$} 
      \label{tab:A}
    
\tablehead{
    \colhead{Bow Shock}            &
    \colhead{$\alpha$$_{2000}$} & 
    \colhead{$\delta$$_{2000}$} &
    \colhead{Source}			
    }
      
\startdata
OS   & 5:43:39.9    &   --01:08:20  & UKIRT  \\       
ON   & 5:44:00.8    &   --00:57:42  & UKIRT  \\            
ON2  & 5:44:06.6    &   --00:53:40  & Spitzer  \\             
\enddata
\vspace{-0.8cm}
\tablecomments{ {\em{a}:} Positions refer to the approximate geometric center.}
\end{deluxetable}

\vspace{3cm}

\begin{deluxetable}{lrrrr}
  \tablewidth{0pt}
  
\centering

\tablecaption{Proper Motions$^a$} 
      \label{tab:B}
    
\tablehead{
    \colhead{Object$^b$}      &
    \colhead{$\Delta$X$^c$} & 
    \colhead{$\Delta$Y$^c$} &
    \colhead{V(X)$^d$}          &
    \colhead{V(Y)$^d$}          			
    }
      
\startdata
NB3  &  -1.2  &  86.6  &   -27.  &   157. \\
NB3   &  7.3  &  70.8   &  -28.  &   186. \\
NB1/2 &  1.3  &  40.3  &   -61.  &   236. \\
NK7  &   0.3  &  24.1    &   0.   &  147.  \\
NK1  &   0.0  &   6.8   &  -48.   &  129.  \\
SK1   &  0.0  &  -6.8   &  -35.  &  -138.  \\
SB1/2  & 0.5  & -38.7  &   -27.  &  -273. \\
SB3a   &  7.7  &  -63.5  &    38.  &  -161. \\
SB3b   &  1.8  &  -79.5   &   20.   &  -143. \\
SB4  &  -3.3  & -127.5  &    13. &   -137. \\
OSa   & 11.3  & -338.5   &    68.  &   -132. \\
OSb   & 29.5  & -362.8    &   47.   &  -150. \\
OSc   & 23.8  & -373.7    &   31.   &  -129. \\
OSd   &  27.8  & -400.4    &  -55.   &  -312. \\
\enddata
\tablecomments{ {\em{a}}: Errors in individual velocities are $\pm$45~km/sec.  {\em{b}}: Knot identifications starting with S or K are from Lee et al. (2007), and starting with O are from this paper.  {\em{c}}: The x,y coordinates are measured in arcsecs from the source along (Y) and across (X) the outflow axis (positive Y to the N and positive X to the W.)  {\em{d}}: Velocities are in km/sec assuming a distance of 400 pc.}
\end{deluxetable}


\begin{thebibliography}{}

\bibitem[Anderson \& Francis(2011)]{Anderson2011} Anthony-Twarog, B.J. 1982, AJ, 87, 1213

\bibitem[Anderson \& Francis(2011)]{Anderson2011} Bally, J. 2016, Ann. Rev. Astron. Astrophys., 54, 491


\bibitem[Anderson \& Francis(2011)]{Anderson2011} Bianchi, E., Codella, C., Ceccarelli, C. et al. 2017, A\&A, 606, L7



\bibitem[Anderson \& Francis(2011)]{Anderson2011} Cabrit, S., Codella, C., Gueth, F., Nisini, B., Gusdorf, A., 
Dougados, C., \& Bacciotti, F. 2007, A\&A, 468, L29

\bibitem[Anderson \& Francis(2011)]{Anderson2011} Cabrit, S., Codella, C., Gueth, F., Gusdorf, A. 2012,
  A\&A, 548, L2

\bibitem[Anderson \& Francis(2011)]{Anderson2011} Casali M., Adamson A., Alves de Oliveira C., et al., 2007, A\&A, 467, 777

\bibitem[Anderson \& Francis(2011)]{Anderson2011} Chen, X., Arce, H.G., Zhang, Q. et al. 2013, ApJ, 768:A110

\bibitem[Anderson \& Francis(2011)]{Anderson2011} Chini, R., Reipurth, B., Sievers, A., Ward-Thompson, D., Haslam, C.G.T., Kreysa, E., \& Lemke, R. 1997, A\&A, 325, 542

\bibitem[Anderson \& Francis(2011)]{Anderson2011} Claussen, M.J., Marvel, K.B., Wootten, A., \& Wilking,
  B.A. 1998, ApJ, 507, L79

\bibitem[Anderson \& Francis(2011)]{Anderson2011} Codella, C., Cabrit, S., Cesaroni, R., Bacciotti, F.,
  Lefloch, B., \& McCaughrean, M. 2007, A\&A, 462, L53

\bibitem[Anderson \& Francis(2011)]{Anderson2011} Codella, C., Cabrit, S., Gueth, F. et al. 2014, A\&A, 568:L5 

\bibitem[Anderson \& Francis(2011)]{Anderson2011} Codella, C., Ceccarelli, C., Cabrit, S. et al. 2016, A\&A, 586, L3

\bibitem[Anderson \& Francis(2011)]{Anderson2011} Connelley, M.S., Reipurth, B., \& Tokunaga, A.T. 2008,
  AJ, 135, 2496

\bibitem[Anderson \& Francis(2011)]{Anderson2011} Correia, S., Zinnecker, H., Ridgway, S.T., \&
  McCaughrean, M.J. 2009, A\&A, 505, 673

\bibitem[Anderson \& Francis(2011)]{Anderson2011} Cyganowski, C.J., Whitney, B.A., Holden, E. et al. 2008, AJ, 136, 2391

\bibitem[Anderson \& Francis(2011)]{Anderson2011} Davis, C.J., Berndsen, A., Smith, M.D., Chrysostomou, A.
  \& Hobson, J. 2000, MNRAS, 314, 241

\bibitem[Anderson \& Francis(2011)]{Anderson2011} Davis, C.J., Gell, R., Khanzadyan, T., Smith, M.D.,
  Jenness, T. 2010, A\&A, 511, 24

\bibitem[Anderson \& Francis(2011)]{Anderson2011} Devine, D., Bally, J., Reipurth, B., Heathcote, S. 1997, AJ, 114, 2095


\bibitem[Anderson \& Francis(2011)]{Anderson2011} Elias, J.H., Joyce, R.R., Liang, M. et al. 2006, SPIE,
  6269, 62694C

\bibitem[Anderson \& Francis(2011)]{Anderson2011} Fitzpatrick, E.L. 1999, PASP, 111, 63

\bibitem[Anderson \& Francis(2011)]{Anderson2011} Frank, A., Ray, T.P., Cabrit, S. et al. 2014, in {\em Protostars and Planets VI}, eds. H. Beuther et al., Univ, of Arizona Press, p. 541

\bibitem[Anderson \& Francis(2011)]{Anderson2011} Galv\'an-Madrid, R., Avila, R., \& Rodr\'\i guez, L.F.
  2004, Rev. Mex. Astron. Astrofis. 40, 31

\bibitem[Anderson \& Francis(2011)]{Anderson2011} Goodwin, S.P., Kroupa, P., Goodman, A., Burkert, A. 2007,
  in Protostars and Planets V, ed. B. Reipurth, D. Jewitt, \& K. Keil
  (Tucson, AZ; Univ. Arizona Press), 133

\bibitem[Anderson \& Francis(2011)]{Anderson2011} Gutermuth, R.A., Myers, P.C., Megeath, S.T. et al. 2008,
  ApJ, 674, 336

\bibitem[Anderson \& Francis(2011)]{Anderson2011} Hodapp, K.W., Walker, C.H., Reipurth, B. et al. 2004,
  ApJ, 601, L79

\bibitem[Anderson \& Francis(2011)]{Anderson2011} Irwin M. J., Lewis J., Hodgkin S., et al., 2004, in {\em
Optimizing Scientific Return for Astronomy through Information
Technologies}, eds. P. J. Quinn \& A. Bridger, Proc. SPIE, 5493, 411

\bibitem[Anderson \& Francis(2011)]{Anderson2011} Koenig, X.P., Leisawitz, D.T., Benford, D.J. et al. 2012, ApJ, 744, 130

\bibitem[Anderson \& Francis(2011)]{Anderson2011} Kounkel, M., Hartmann, L., Loinard, L. et al. 2017, ApJ, 834, A142

\bibitem[Anderson \& Francis(2011)]{Anderson2011} Krisciunas, K., Sinton, W., Tholen, K. et al. 1987, PASP, 99,887

\bibitem[Anderson \& Francis(2011)]{Anderson2011} Larson, R.B. 1972, MNRAS, 156, 437 

\bibitem[Anderson \& Francis(2011)]{Anderson2011} Lee, C.-F., Mundy, L.G., Reipurth, B., Ostriker, E.C., \&
  Stone, J.M. 2000, ApJ, 542, 925

\bibitem[Anderson \& Francis(2011)]{Anderson2011} Lee, C.-F., Ho, P.T.P., Beuther, H., Bourke, T.L., Zhang,
  Q, Hirano, N., \& Shang, H. 2006, ApJ, 639, 292

\bibitem[Anderson \& Francis(2011)]{Anderson2011} Lee, C.-F., Ho, P.T.P., Hirano, N., Beuther, H., Bourke, T.L., Shang, H., \& Zhang, Q. 2007, ApJ, 659, 499


\bibitem[Anderson \& Francis(2011)]{Anderson2011} Lee, C.-F., Ho, P.T.P., Bourke, T.L., Hirano, N., Shang,
  H., \& Zhang, Q. 2008, ApJ, 685, 1026

\bibitem[Anderson \& Francis(2011)]{Anderson2011} Lee, C.-F., Hirano, N., Zhang, Q., Shang, H., Ho, P., Krasnopolsky, R. 2014, ApJ, 786:A114

\bibitem[Anderson \& Francis(2011)]{Anderson2011} Lee, C.-F., Hirano, N., Zhang, Q., Shang, H., Ho, P.T.P.,
  Mizuno, Y. 2015, ApJ, 805: A186

\bibitem[Anderson \& Francis(2011)]{Anderson2011} Lee, C.-F., Li, Z.-Y., Ho, P.T.P. et al. 2017a, ApJ, 843, A27


\bibitem[Anderson \& Francis(2011)]{Anderson2011} Lee, C.-F., Li, Z.-Y., Ho, P.T.P. et al. 2017b, Science Advances, 3:e1602935

\bibitem[Anderson \& Francis(2011)]{Anderson2011} Lee, C.-F., Li, Z.-Y.,  Ching, T.-C. et al. 2018a, ApJ, 854, A56

\bibitem[Anderson \& Francis(2011)]{Anderson2011} Lee, C.-F., Li, Z.-Y., Codella, C. et al. 2018b, ApJ, 856, A14

\bibitem[Anderson \& Francis(2011)]{Anderson2011} Leggett, S.K., Currie, M.J., Varricatt, W.P. et al. 2006,
  MNRAS, 373, 781

\bibitem[Anderson \& Francis(2011)]{Anderson2011} Leurini, S., Codella, C., Cabrit, S. et al. 2016, A\&A, 595, L4

\bibitem[Anderson \& Francis(2011)]{Anderson2011} Megeath, S.T., Gutermuth, R., Muzerolle, J. et al. 2012,
  AJ, 144:A192


\bibitem[Anderson \& Francis(2011)]{Anderson2011} Miettinen, O. 2012, A\&A, 545:A3


\bibitem[Anderson \& Francis(2011)]{Anderson2011} Podio, L., Codella, C., Gueth, F. et al. 2015, A\&A, 581, A85

\bibitem[Anderson \& Francis(2011)]{Anderson2011} Raga, A.C., Cant\'o, J., Binette, L., Calvet. N. 1990, ApJ, 364, 601

\bibitem[Anderson \& Francis(2011)]{Anderson2011} Raga, A.C., Reipurth, B., Esquivel, A. et al. 2017, Rev. Mex. Astron. Astrofis., 53, 485

\bibitem[Anderson \& Francis(2011)]{Anderson2011} Rayner, J.T., Cushing, M.C., Vacca, W.D. 2009, ApJS, 185, 289

\bibitem[Anderson \& Francis(2011)]{Anderson2011} Reipurth, B. 2000, AJ, 120, 3177

\bibitem[Anderson \& Francis(2011)]{Anderson2011} Reipurth, B. \& Bally, J. 2001,
Ann. Rev. Astron. Astrophys., 39, 403


\bibitem[Anderson \& Francis(2011)]{Anderson2011} Reipurth, B. \& Clarke, C.J. 2001, AJ, 122, 432

\bibitem[Anderson \& Francis(2011)]{Anderson2011} Reipurth, B. \& Mikkola, S. 2012, Nature, 492, 221

\bibitem[Anderson \& Francis(2011)]{Anderson2011} Reipurth, B., Bally, J., Devine, D. 1997, AJ, 114, 2708

\bibitem[Anderson \& Francis(2011)]{Anderson2011} Reipurth, B., Heathcote, S., Yu, K.C. et al. 2000, ApJ, 534, 317

\bibitem[Anderson \& Francis(2011)]{Anderson2011} Reipurth, B., Heathcote, S., Morse, J. et al. 2002, AJ, 123, 362 

\bibitem[Anderson \& Francis(2011)]{Anderson2011} Reipurth, B., Mikkola, S., Connelley, M., \& Valtonen, M.
  2010, ApJ, 725, L56

\bibitem[Anderson \& Francis(2011)]{Anderson2011} Reipurth, B., Clarke, C.J., Boss, A.P., Goodwin, S.P.,
  Rodriguez, L.F., Stassun, K.G., Tokovinin, A., \& Zinnecker, H.
  2014, in Protostars and Planets VI, eds. H. Beuther et al., (Tucson,
  AZ; Univ. Arizona Press), p.267

\bibitem[Anderson \& Francis(2011)]{Anderson2011} Sahu, D., Minh, Y.C., Lee, C.-F. et al. 2018, MNRAS, 475, 5322

\bibitem[Anderson \& Francis(2011)]{Anderson2011} Smith, M.D., O'Connell, B., \& Davis, C.J. 2007, A\&A, 466, 565

\bibitem[Anderson \& Francis(2011)]{Anderson2011} Tabone, B., Cabrit, S., Bianchi, E. et al. 2017, A\&A, 607, L6

\bibitem[Anderson \& Francis(2011)]{Anderson2011} Takami, M., Takakuwa, S., Momose, M., Hayashi, M., Davis,
  C.J., Pyo, T.-S., Nishikawa, T., \& Kohno, K. 2006, PASJ, 58, 563

\bibitem[Anderson \& Francis(2011)]{Anderson2011} Takami, M., Karr, J.L., Koh, H., Chen, H.-H., \& Lee, H.-T. 2010,
ApJ 720, 2010

\bibitem[Anderson \& Francis(2011)]{Anderson2011} Tokunaga, A.T., Kobayashi, N., Bell, J. et al. 1998, in SPIE Conf Ser., ed. A.M. Fowler, vol. 3354, 512

\bibitem[Anderson \& Francis(2011)]{Anderson2011} Tokunaga, A.T., Simons, D.A., \& Vacca, W.D. 2002, PASP, 114, 180

\bibitem[Anderson \& Francis(2011)]{Anderson2011} Umbreit, S., Spurzem, R., Henning, Th., Klahr, H.,
  Mikkola, S. 2011, ApJ, 743:A106

\bibitem[Anderson \& Francis(2011)]{Anderson2011} Valtonen, M.J. \& Mikkola, S. 1991, ARA\&A, 29, 9

\bibitem[Anderson \& Francis(2011)]{Anderson2011} van Belle, G.T., Lane, B.F., Thompson, R.R. et al. 1999,
AJ, 117, 521

\bibitem[Anderson \& Francis(2011)]{Anderson2011} Wiseman, J., Wootten, A., Zinnecker, H., \& McCaughrean,
  M. 2001, ApJ, 550, L87

\bibitem[Anderson \& Francis(2011)]{Anderson2011} Zinnecker, H., Bastien, P., Arcoragi, J.-P., Yorke, H. W.
  1992, A\&A, 265, 726

\bibitem[Anderson \& Francis(2011)]{Anderson2011} Zinnecker, H., McCaughrean, M., \& Rayner, J. 1998,
  Nature, 394, 862
\end{thebibliography}
\end{document}